\documentclass[cameraready]{Interspeech}
\title{Time–Frequency Weighted Losses for Phoneme Reconstruction \\in DNN-Based Speech Enhancement}
\author[orcid=0009-0006-7531-2051]{Nasser-Eddine}{Monir}
\author[orcid=0000-0002-8561-0961]{Paul}{Magron}
\author[orcid=0000-0002-6848-0114,]{Romain}{Serizel}
\address{Universit\'{e} de Lorraine, CNRS, Inria, LORIA, F-54000 Nancy, France}

\email{\{nasser-eddine.monir, paul.magron\}@inria.fr, romain.serizel@loria.fr}


\usepackage{comment}
\usepackage{multirow}
\usepackage{pgfplots}
\pgfplotsset{compat=1.18}
\usepgfplotslibrary{groupplots}

\begin{document}

\maketitle

\begin{abstract}
Conventional training losses for speech enhancement based on the signal-to-distortion ratio (SDR) treat all time–frequency (TF) regions uniformly, overlooking the fine-grained spectral cues that are relevant to specific phoneme intelligibility. We propose a TF weighting framework that modulates the SDR objective based on local speech presence, speech-to-interference ratio (SIR), and spectral flux. By integrating these factors into a differentiable objective, the framework emphasizes TF bins with high speech–noise competition while also accounting for transient cues such as consonant bursts. Experimental results show that our approach improves objective frequency-weighted enhancement metrics, as well as phoneme recognition accuracy, particularly for consonants. Spectral analysis shows better reconstruction of mid-frequency structures at less adverse SIR.
\end{abstract} 

\section{Introduction}

Deep learning–based multichannel speech enhancement (MCSE) has evolved toward end-to-end neural-spatial frameworks that jointly model spectral and inter-channel dependencies~\cite{luo2020end, lee2021inter, quan2024spatialnet}. State-of-the-art models leverage temporal convolutional networks and self-attention to optimize spatial filtering.
%These architectures prioritize global reconstruction and may still overlook fine-grained phonetic cues.  %% j'enlève car tu créé de la confusion entre l'impact des architectures et celui des loss
MCSE algorithms are commonly trained using signal-to-distortion ratio (SDR)-based objectives~\cite{luo2020end}. While $\mathrm{SDR}$-based losses are effective for improving overall signal reconstruction and interference reduction, they implicitly treat all time or time–frequency (TF) regions uniformly. By assigning equal importance to all TF regions, SDR-based losses overlook the uneven perceptual contribution of different speech components, especially in noisy conditions.

%This assumption neglects the fact that perceptual relevance is unevenly distributed across the speech signal, particularly in noisy conditions.

From a phonetic standpoint, intelligibility depends critically on specific acoustic cues that are localized in both time and frequency~\cite{Miller1955, Stevens1998}. Consonantal transitions, plosive bursts, fricative noise bands, and formant trajectories carry disproportionate linguistic information compared to steady-state vowel regions~\cite{Klatt1976, Fant1960}. Moreover, masking effects are the strongest when speech and noise magnitudes are comparable, leading to degraded transmission of these cues~\cite{Gelfand1985, Phatak2008}. In hearing-assistive scenarios with competing sound sources or background noise, preserving transient and mid-frequency speech structures is particularly important~\cite{Bronkhorst2000, Kates2014, VanDenBogaert2009}. Consequently, training objectives that emphasize regions of strong speech–noise interaction may better reflect perceptual priorities than globally uniform losses~\cite{Greenberg1993, MaHuLoizou2009, Monir2025FreqWeightedSDR}. This intuition is also related to glimpsing theories of speech perception, which emphasize the importance of local spectro-temporal regions that remain informative in the presence of masking noise~\cite{Cooke2006}.

Monir et al.~\cite{Monir2025FreqWeightedSDR} previously introduced frequency-weighted SDR formulations to improve phoneme-level behavior. These approaches incorporated perceptual band-importance functions and local signal-to-interference (SIR) modulation, yet did not explicitly combine speech presence, speech–noise competition, and transient dynamics within a unified weighting framework. %model the joint interaction between speech presence, speech–noise competition, and transient dynamics.

% In this work, we propose a TF weighting strategy integrating speech presence, local $\mathrm{SIR}$, and spectral flux into a differentiable $\mathrm{SDR}$ objective. The framework emphasizes TF bins with strong speech–noise competition while remaining adaptive to transient phonetic cues. We evaluate the proposed formulations using utterance-level separation metrics, intelligibility measures, word recognition error rate (WER), and phoneme accuracy (PA), with particular focus on consonants and plosives. Spectral analyses further illustrate how the proposed weighting reshapes reconstruction behavior across noise conditions.
In this work, we investigate a structured TF-weighting framework grounded in speech–noise competition. We first model the speech-noise competition explicitly through SIR and speech-presence gating. We then extend this formulation to transient phonetic cues via spectral flux. Finally, we examine a data-driven alternative where spectral weights are fully learned. This progression allows us to analyze the respective roles of perceptual inductive bias and learned weighting in phoneme-level enhancement. %Our experiments show the potential of this approach in terms of enhancement metrics and phoneme recognition. Besides, spectral analysis reveals better reconstruction of mid-frequency structures that are critical for intelligibility.

\section{Methodology}

\subsection{TF-weighted SDR}

Conventional $\mathrm{SDR}$-based objectives treat all TF bins uniformly, although their perceptual relevance strongly depends on local speech--noise interaction~\cite{Gelfand1985, Greenberg1993}. In particular, intelligibility is mainly affected in regions where speech and noise have comparable magnitudes, leading to strong competition and masking~\cite{Gelfand1985}. In contrast, bins dominated by clean speech or containing negligible energy contribute less to perceptual errors~\cite{Greenberg1993}. Therefore, we introduce a TF-weighted $\mathrm{SDR}$ framework that explicitly models local speech--noise competition.

Let $S(f,t)$, $N(f,t)$, and $\widehat{S}(f,t)$ denote the Mel-band magnitudes of the clean speech, noise, and estimated speech signals in the TF domain, respectively. In practice we compute their short-time Fourier transform and group frequencies using a Mel scale. Let $S_{\mathrm{proj}}(f,t)$ denote the orthogonal projection of $\widehat{S}(f,t)$ onto $S(f,t)$, and $E_{\mathrm{dist}}(f,t) = \widehat{S}(f,t) - S_{\mathrm{proj}}(f,t)$ the residual distortion component. We then define the TF-weighted SDR loss is as follows:
\begin{equation}
\mathcal{L}_{w}
=
-10 \log_{10}
\frac{\sum_{f,t} w(f,t)\,|S_{\mathrm{proj}}(f,t)|^2}
{\sum_{f,t} w(f,t)\,|E_{\mathrm{dist}}(f,t)|^2},
\end{equation}
where $w(f,t)\ge0$ controls the contribution of each TF bin. 
When $w(f,t)=1$, this formulation reduces to an unweighted TF-domain SDR.
%While distinct from the time-domain $\mathrm{SDR}$ loss~\cite{LeRoux2019} used as baseline ($\mathcal{L}_{\mathrm{T}}$), it relies on the same projection and decomposition}.
Note that it is distinct from a time-domain $\mathrm{SDR}$ loss~\cite{LeRoux2019}, which is used here as baseline and denoted $\mathcal{L}_{\mathrm{T}}$.
Monir et al.~\cite{Monir2025FreqWeightedSDR} proposed several weighting strategies based on the local SIR defined as $\mathrm{SIR}(f,t)=10\log_{10}\frac{|S(f,t)|^2}{|N(f,t)|^2}$. Among these, the best performing weights are $w(f,t)=\mathrm{softmax}(-\log(\mathrm{SIR}(f,t)))$, which yield a loss denoted $\mathcal{L}_{\mathrm{logSIR}}$. While promising, these approaches do not fully account for the speech–noise competition.

\begin{table*}[t]
\centering
\caption{Utterance-level performance for the white noise (WN) and speech-shaped noise (SSN) test sets averaged across input $\mathrm{SIR}$ levels between $\mathrm{-8}$~dB and $\mathrm{8}$~dB. Note that input (FW)-SDR and (FW)-SAR are not reported as they are infinite.}
\vspace{-1em}
\label{tab:global_results_full}
\setlength{\tabcolsep}{3pt}
\resizebox{\textwidth}{!}{
\begin{tabular}{l|ccccccc|ccccccc}
\toprule
\multirow{2}{*}{Loss} 
& \multicolumn{7}{c}{WN}
& \multicolumn{7}{c}{SSN} \\
\cmidrule(lr){2-8}
\cmidrule(lr){9-15}
& $\mathrm{SIR}$
& $\mathrm{SAR}$
& $\mathrm{SDR}$
& $\mathrm{FW\text{-}SIR}$
& $\mathrm{FW\text{-}SAR}$
& $\mathrm{FW\text{-}SDR}$
& $\mathrm{STOI}$
& $\mathrm{SIR}$
& $\mathrm{SAR}$
& $\mathrm{SDR}$
& $\mathrm{FW\text{-}SIR}$
& $\mathrm{FW\text{-}SAR}$
& $\mathrm{FW\text{-}SDR}$
& $\mathrm{STOI}$
\\
\midrule

% Input
Input
& -2.0 & -- & -- & -6.0 & -- & -- & 0.50
& -2.0 & -- & -- & -4.8 & -- & -- & 0.54 \\
\midrule

% Baseline
$\mathcal{L}_{\mathrm{T}}$
& 13.3 & \textbf{2.4} & \textbf{1.7} & 5.1 & \textbf{3.5} & 4.7 & 0.63
& 14.3 & \textbf{1.2} & \textbf{0.7} & 7.5 & 2.2 & 4.4 & \textbf{0.63} \\

% L_log_softmax_SIR
$\mathcal{L}_{\mathrm{logSIR}}$
& 16.2 & 1.2 & 0.9 & 7.6 & 3.1 & 4.8 & 0.61
& 14.2 & -1.1 & -1.6 & 7.5 & 2.6 & 3.9 & 0.57 \\

% L\_LEARN\_SOFT\_INIT\_ANSI
$\mathcal{L}_{\mathrm{learn}}$
& 16.1 & 1.9 & 1.6 & 7.5 & 3.4 & 5.0 & \textbf{0.64}
& 15.0 & 0.3 & -0.1 & 8.2 & 2.2 & 4.6 & 0.59 \\

% SigSIR\_SigSP\_LEARNABLE\_TAUS
$\mathcal{L}_{\mathrm{SIR}\cdot\mathrm{SP}}$
& 15.1 & 2.0 & 1.5 & 6.6 & 3.3 & \textbf{5.2} & 0.63
& 13.4 & 0.4 & -0.2 & 6.8 & 1.8 & 3.6 & 0.57 \\

% SIR\_S\_SFlux\_a0.2
$\mathcal{L}_{\mathrm{SIR}\cdot\mathrm{SP}\cdot\mathrm{SF}}$
& \textbf{17.2} & 1.8 & 1.5 & \textbf{8.4} & 2.8 & \textbf{5.2} & \textbf{0.64}
& \textbf{15.1} & 0.8 & 0.4 & \textbf{8.3} & \textbf{2.3} & \textbf{4.9} & 0.61 \\

\bottomrule
\end{tabular}
}
\end{table*}
\subsection{Proposed weighting schemes}

Intuitively, one desires high weight values when both $|S(f,t)|$ and $|N(f,t)|$ are large, as these bins correspond to strong speech--noise competition. Conversely, bins with dominant speech (high local $\mathrm{SIR}$) or very low speech energy should receive smaller weights in order to avoid over-optimizing perceptually less critical regions. We propose the following schemes that comply with these requirements. The first two schemes are grounded in explicit modeling of speech--noise competition, while the third examines whether a purely learned spectral profile can achieve similar benefits.\\

\noindent \textbf{SIR--speech presence weighting} First, let us define the two following gating functions:
\begin{align}
\label{gate_sir}
g_{\mathrm{SIR}}(f,t) &=\sigma(-\mathrm{SIR}(f,t)+\tau_1), \\
g_{\mathrm{SP}}(f,t)&=\sigma(|S(f,t)|^\gamma+\tau_2),
\end{align}
where $\sigma(\cdot)$ denotes the sigmoid function, and $\tau_1$ and $\tau_2$ are learned thresholds. We propose the following weights:
\begin{equation}
w(f,t)
=
g_{\mathrm{SIR}}(f,t)
\cdot
g_{\mathrm{SP}}(f,t)
\end{equation}
which emphasizes TF bins where speech remains active despite low $\mathrm{SIR}$. The corresponding loss is denoted $\mathcal{L}_{\mathrm{SIR}\cdot\mathrm{SP}}$.\\

\noindent \textbf{SIR--speech presence--spectral flux weighting}
When $|S|$ is weak but perceptually relevant, as in several transient phonemes (e.g., plosives), a purely magnitude-based weighting may underestimate their importance. To capture such rapidly varying cues, we consider the frame-wise spectral flux
\begin{equation}
\mathrm{SF}(t)
=
\frac{
\sum_f \max(|S(f,t)| - |S(f,t-1)|, 0)
}{
\sum_f |S(f,t-1)| + \epsilon
},
\end{equation}
which measures the relative increase of spectral energy between consecutive frames. 
We incorporate this quantity into the weighting scheme as follows:
\begin{equation}
w(f,t)
=
g_{\mathrm{SIR}}(f,t)
\cdot
g_{\mathrm{SP}}(f,t)
\cdot
\left(1+k\,\sigma(\mathrm{SF}(t))\right),
\end{equation}
where $k$ is a scaling factor adjusting the relative importance of the spectral flux. This scheme increases sensitivity to transient phonetic structures while preserving speech--noise competition modeling. The corresponding loss is denoted $\mathcal{L}_{\mathrm{SIR}\cdot\mathrm{SP}\cdot\mathrm{SF}}$.\\

\noindent \textbf{Learnable weights} Lastly, we consider the following scheme:
\begin{equation}
w(f)=\mathrm{softmax}(\theta_f),
\end{equation}
where the spectral parameters $\theta_f$ are learned and time-independent. We initialize $\theta_f$ with ANSI~1997 band-importance weights~\cite{Pavlovic2018SII}. The corresponding loss is denoted $\mathcal{L}_{\mathrm{learn}}$.

% All schemes are fully differentiable and retain the $\mathrm{SDR}$ structure while reshaping the optimization toward perceptually relevant TF regions. 
% The resulting loss functions are denoted $\mathcal{L}_{\mathrm{SIR}\cdot\mathrm{SP}}$, 
% $\mathcal{L}_{\mathrm{SIR}\cdot\mathrm{SP}\cdot\mathrm{SF}}$, and 
% $\mathcal{L}_{\mathrm{learn}}$. 
% We also evaluated alternative gating functions (e.g., ReLU-based variants), but for brevity we report the formulations that yielded the most stable performance.

%%%

\section{Experimental protocol}

\subsection{Acoustic conditions and dataset}
We adopt the data generation and acoustic simulation protocol of Monir et al.~\cite{Monir2025FreqWeightedSDR} for training and validation. Clean speech is drawn from LibriSpeech~\cite{Panayotov2015}, and noise consists of SSN and ecological noise from the Disconoise dataset~\cite{Furnon2021}. Reverberant mixtures are generated using room impulse responses (RIRs) simulated with Pyroomacoustics~\cite{Scheibler2018}, with randomized room geometry and $\mathrm{RT}_{60}$, and a 4-channel binaural hearing-aid microphone configuration. Mixtures are created over $\mathrm{SIR}$ values in $[-10,10]$~dB.
For evaluation, we consider two dedicated test sets: one with WN and one with SSN, computed on 10 speakers (5 male, 5 female). These test sets follow a controlled configuration, with the target speech placed at $0^\circ$ (front) and the masker at $45^\circ$ (right), using measured RIRs from the Binaurec dataset~\cite{Delebecque2023} at $\mathrm{SIR}$ levels from -8~dB to 8~dB.

\subsection{Speech enhancement algorithm}
As MCSE algorithm we use FaSNet, an end-to-end time-domain two-stage adaptive beamformer~\cite{luo2020end}. We follow the training process~\footnote{The source code for this work is publicly available at \url{https://github.com/Nasseredd/fw-se-loss}} of Monir et al.~\cite{Monir2025FreqWeightedSDR}, relying on Asteroid~\cite{Pariente2020Asteroid}. For a binaural hearing-aid configuration, microphone permutation is disabled to maintain a fixed front-left reference channel. The factor $k$ in $\mathcal{L}_{\mathrm{SIR}\cdot\mathrm{SP}\cdot\mathrm{SF}}$ is tuned on the validation set to $0.2$.  %Utterances are truncated to 10~s, with a patience of 15 epochs and a batch size of 1~\cite{Anonymous2025FWSDR}.

\subsection{Performance measures}

As objective metrics, we report scale-invariant SDR, SIR, and signal-to-artifact ratios (SAR) computed in the time-domain~\cite{Vincent2006, LeRoux2019} on the left ear, which is the best one~\cite{bronkhorst1988}. In addition, we report their intelligibility-oriented frequency-weighted (FW) counterparts~\cite{Greenberg1993, MaHuLoizou2009}, with weights proportional to $|S(f,t)|^{0.2}$~\cite{MaHuLoizou2009,Monir2025FreqWeightedSDR}. We also include short-time objective intelligibility (STOI)~\cite{Taal2010} to assess intelligibility.

We additionally report the word error rate (WER) and phoneme accuracy (PA) when performing speech and phoneme recognition after MCSE using Wav2Vec2-based models~\cite{Baevski2020Wav2Vec, vitouphy}. PA is defined as the percentage of correctly predicted phoneme tokens within their annotated temporal boundaries. These metrics serve as objective proxies for intelligibility~\cite{spille2018human} and phonetic reconstruction. They do not replace listening tests but provide a reproducible evaluation across conditions. Metrics are computed on the left ear~\cite{bronkhorst1988}. For all metrics except for the WER, higher is better. Statistical significance is assessed via the Wilcoxon signed-rank test~\cite{wilcoxon1945individual} performed between each method against the baseline, at a level of $\alpha = 5\%$. Significant gains are highlighted in bold fonts in the tables.
%and, for brevity, selectively discussed in the text.

\begin{table*}[t]
\centering
\caption{Performance for phoneme categories (consonants and vowels) for the WN and SSN test sets averaged across input $\mathrm{SIR}$ levels.}
\vspace{-1em}
\label{tab:phoneme_results_compact}
\setlength{\tabcolsep}{3pt}
\resizebox{\textwidth}{!}{
\begin{tabular}{l l | ccccccc | ccccccc}
\toprule
\multirow{2}{*}{} & \multirow{2}{*}{Loss}
& \multicolumn{7}{c}{WN}
& \multicolumn{7}{c}{SSN} \\
\cmidrule(lr){3-9} \cmidrule(lr){10-16}
& &
$\mathrm{SIR}$ & $\mathrm{SAR}$ & $\mathrm{SDR}$ &
$\mathrm{FW\text{-}SIR}$ & $\mathrm{FW\text{-}SAR}$ & $\mathrm{FW\text{-}SDR}$ &
$\mathrm{PA}$ &
$\mathrm{SIR}$ & $\mathrm{SAR}$ & $\mathrm{SDR}$ &
$\mathrm{FW\text{-}SIR}$ & $\mathrm{FW\text{-}SAR}$ & $\mathrm{FW\text{-}SDR}$ &
$\mathrm{PA}$ \\
\midrule

% =================== CONSONANT ===================
\multirow{2}{*}{Consonants}
& $\mathcal{L}_{\mathrm{T}}$
& 12.8 & \textbf{2.0} & \textbf{1.4} & 6.5 & \textbf{3.9} & 6.3 & 34.0
& 14.7 & \textbf{0.8} & \textbf{0.5} & 9.9 & \textbf{2.4} & \textbf{6.4} & 43.6 \\

% & $\mathcal{L}_{\mathrm{learn}}$
% & 15.4 & 1.6 & 1.3 & 9.0 & 3.6 & 6.4 & 35.2
% & \textbf{15.8} & 0.1 & -0.2 & \textbf{11.0} & 2.3 & 6.2 & 42.6 \\

% & $\mathcal{L}_{\mathrm{SIR}\cdot\mathrm{SP}}$
% & 13.6 & 1.6 & 1.1 & 7.3 & 3.6 & 6.5 & 35.7
% & 12.2 & 0.2 & -0.6 & 7.5 & 1.9 & 4.6 & 43.1 \\

& $\mathcal{L}_{\mathrm{SIR}\cdot\mathrm{SP}\cdot\mathrm{SF}}$
& \textbf{16.1} & 1.4 & 1.1 & \textbf{9.7} & 3.0 & \textbf{6.7} & \textbf{36.1}
& 14.6 & 0.4 & 0.0 & 9.8 & 2.1 & 6.1 & \textbf{45.5} \\

\midrule

% =================== VOWEL ===================
\multirow{2}{*}{Vowels}
& $\mathcal{L}_{\mathrm{T}}$
& 14.4 & \textbf{3.2} & \textbf{2.7} & 11.4 & \textbf{4.6} & \textbf{8.4} & 43.7
& 13.8 & \textbf{1.9} & \textbf{1.3} & 12.5 & 3.0 & 7.6 & 46.4 \\

% & $\mathcal{L}_{\mathrm{learn}}$
% & 16.8 & 2.6 & 2.4 & 13.9 & 4.3 & 7.7 & 44.3
% & 14.8 & 0.5 & 0.3 & 13.5 & 3.0 & 7.7 & 44.2 \\

% & $\mathcal{L}_{\mathrm{SIR}\cdot\mathrm{SP}}$
% & 15.8 & 2.8 & 2.4 & 12.9 & 4.4 & 8.0 & 44.8
% & 13.9 & 1.1 & 0.6 & 12.6 & 3.0 & 7.5 & 45.8 \\

& $\mathcal{L}_{\mathrm{SIR}\cdot\mathrm{SP}\cdot\mathrm{SF}}$
& \textbf{17.8} & 2.7 & 2.5 & \textbf{14.8} & 3.9 & 7.7 & \textbf{45.5}
& \textbf{15.1} & 1.2 & 0.8 & \textbf{13.8} & \textbf{3.3} & \textbf{8.4} & \textbf{47.9} \\

\bottomrule
\end{tabular}
}
\end{table*}
\definecolor{myblue}{HTML}{0057B8}   % Best Model
\definecolor{darkred}{HTML}{B22222}  % Intermediate 1
\definecolor{darkgray}{HTML}{555555} % Intermediate 2
\definecolor{mypurple}{HTML}{7F22FE} % Intermediate 2

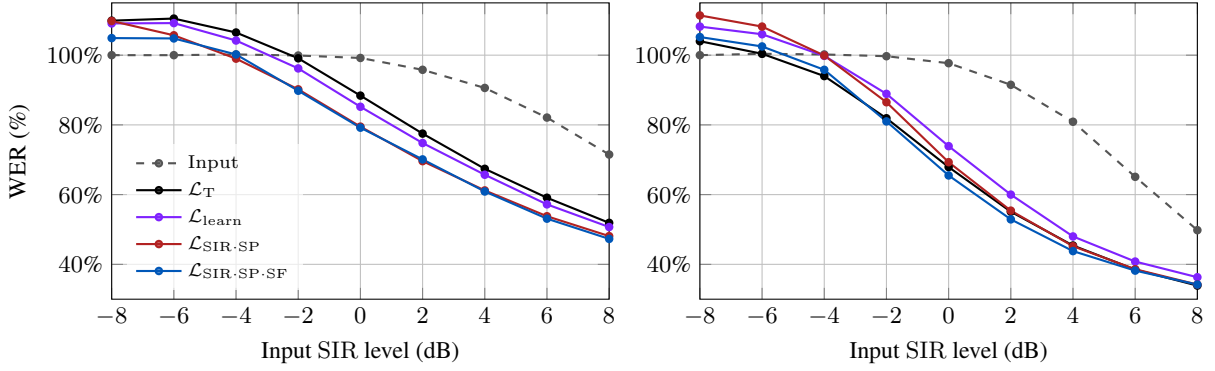
\begin{figure*}[t]
\centering
\begin{tikzpicture}

\begin{groupplot}[
    group style={
        group size=2 by 1,
        horizontal sep=1.2cm
    },
    width=0.48\textwidth,
    height=5.5cm, 
    xlabel={Input $\mathrm{SIR}$ level (dB)},
    ymin=30, ymax=115, 
    xmin=-8, xmax=8,
    xtick={-8,-6,-4,-2,0,2,4,6,8},
    yticklabel={\pgfmathprintnumber{\tick}\%}, 
    grid=major
]

% ================= PLOT 1 (WN) =================
\nextgroupplot[
    ylabel={WER (\%)},
    % Legend configuration inside the first plot
    legend style={
        at={(0.03,0.03)},     % Positioning: 3% from left, 3% from bottom
        anchor=south west, 
        draw=none,           % Removes the black border line
        fill=white,          % Fills background to keep text legible over grid
        fill opacity=0.7,    % Optional: makes background slightly transparent
        text opacity=1,
        font=\scriptsize,    % Smaller font to fit 5 entries inside the box
        cells={anchor=west}
    }
]

\addplot[darkgray,dashed,thick,mark=*,mark size=1.2pt, mark options={solid}] coordinates {
(-8,100.0) (-6,100.0) (-4,100.2) (-2,99.9) (0,99.2) (2,95.8) (4,90.6) (6,82.1) (8,71.5)
};
\addlegendentry{${\mathrm{Input}}$}

\addplot[black,thick,mark=*,mark size=1.2pt] coordinates {
(-8,109.9) (-6,110.5) (-4,106.5) (-2,99.1) (0,88.4) (2,77.5) (4,67.4) (6,59.1) (8,51.9)
};

\addlegendentry{$\mathcal{L}_{\mathrm{T}}$}

\addplot[mypurple,thick,mark=*,mark size=1.2pt] coordinates {
(-8,109.1) (-6,109.2) (-4,104.2) (-2,96.2) (0,85.2) (2,74.8) (4,65.7) (6,57.2) (8,50.7)
};
\addlegendentry{$\mathcal{L}_{\mathrm{learn}}$}

\addplot[darkred,thick,mark=*,mark size=1.2pt] coordinates {
(-8,109.8) (-6,105.7) (-4,99.0) (-2,90.2) (0,79.5) (2,69.6) (4,61.2) (6,53.8) (8,48.1)
};
\addlegendentry{$\mathcal{L}_{\mathrm{SIR}\cdot\mathrm{SP}}$}

\addplot[myblue,thick,mark=*,mark size=1.2pt] coordinates {
(-8,104.9) (-6,104.8) (-4,100.2) (-2,89.8) (0,79.2) (2,70.1) (4,60.9) (6,53.1) (8,47.3)
};
\addlegendentry{$\mathcal{L}_{\mathrm{SIR}\cdot\mathrm{SP}\cdot\mathrm{SF}}$}

% ================= PLOT 2 (SSN) =================
\nextgroupplot[
    ylabel={} 
]

\addplot[darkgray,dashed,thick,mark=*,mark size=1.2pt, mark options={solid}] coordinates {
(-8,100.0) (-6,100.4) (-4,100.2) (-2,99.7) (0,97.7) (2,91.5) (4,80.9) (6,65.1) (8,49.8)
};

\addplot[black,thick,mark=*,mark size=1.2pt] coordinates {
(-8,104.0) (-6,100.4) (-4,94.0) (-2,81.9) (0,67.9) (2,55.1) (4,45.4) (6,38.5) (8,33.9)
};

\addplot[mypurple,thick,mark=*,mark size=1.2pt] coordinates {
(-8,108.2) (-6,106.0) (-4,99.8) (-2,88.9) (0,73.9) (2,60.0) (4,48.0) (6,40.8) (8,36.3)
};

\addplot[darkred,thick,mark=*,mark size=1.2pt] coordinates {
(-8,111.4) (-6,108.2) (-4,99.9) (-2,86.5) (0,69.3) (2,55.4) (4,45.2) (6,38.6) (8,34.2)
};

\addplot[myblue,thick,mark=*,mark size=1.2pt] coordinates {
(-8,105.2) (-6,102.5) (-4,95.8) (-2,81.0) (0,65.5) (2,52.9) (4,43.8) (6,38.2) (8,34.2)
};

\end{groupplot}

\end{tikzpicture}
\caption{$\mathrm{WER}$ (\%) as a function of the input $\mathrm{SIR}$ level for WN (left) and SSN (right) test sets.}
\label{fig:wer_snr}
\end{figure*}

\section{Results}

\subsection{Results at the utterance level}

% Table~\ref{tab:global_results_full} displays utterance-level performance in terms of classical, frequency-weighted, and intelligibility metrics. First, we observe that under $\mathrm{WN}$, all losses improve $\mathrm{SIR}$ compared to the baseline. The frequency-learned loss $\mathcal{L}_{\mathrm{learn}}$ and the time–frequency variant $\mathcal{L}_{\mathrm{SIR\cdot SP}}$ provide clear gains in $\mathrm{SIR}$ and $\mathrm{FW\text{-}SIR}$, while the~$\mathcal{L}_{\mathrm{SIR\cdot SP\cdot SF}}$ achieves the strongest overall improvement in both metrics. Introducing the $\mathrm{SIR}$-based weighting or learning frequency weights does not lead to a substantial degradation in $\mathrm{SAR}$, $\mathrm{FW\text{-}SAR}$, $\mathrm{SDR}$, $\mathrm{FW\text{-}SDR}$, or $\mathrm{STOI}$, as results remain overall comparable across losses, with only slight variations relative to the baseline.
Table~\ref{tab:global_results_full} displays utterance-level performance in terms of classical, frequency-weighted, and intelligibility metrics. First, we observe that under $\mathrm{WN}$, all losses yield an improved $\mathrm{SIR}$ and $\mathrm{FW\text{-}SIR}$, with the~$\mathcal{L}_{\mathrm{SIR\cdot SP\cdot SF}}$ achieving the strongest overall improvement. Introducing the $\mathrm{SIR}$-based weighting or learning frequency weights does not lead to a substantial degradation in $\mathrm{SAR}$, $\mathrm{FW\text{-}SAR}$, $\mathrm{SDR}$, $\mathrm{FW\text{-}SDR}$, or $\mathrm{STOI}$, as results remain overall comparable across losses, with only slight variations relative to the baseline.

Under $\mathrm{SSN}$, both $\mathcal{L}_{\mathrm{learn}}$ and $\mathcal{L}_{\mathrm{SIR\cdot SP\cdot SF}}$ slightly improve interference reduction compared to the baseline. In contrast, $\mathcal{L}_{\mathrm{logSIR}}$ and $\mathcal{L}_{\mathrm{SIR\cdot SP}}$ do not consistently reduce interference. %and may slightly reduce it. 
On the other hand, $\mathrm{SAR}$ and $\mathrm{SDR}$ slightly decrease for the weighted losses, with a more pronounced reduction for $\mathcal{L}_{\mathrm{logSIR}}$ and a smaller one for $\mathcal{L}_{\mathrm{SIR\cdot SP\cdot SF}}$. Frequency-weighted metrics remain overall comparable to the baseline, with a more noticeable drop for $\mathcal{L}_{\mathrm{SIR\cdot SP}}$, whereas $\mathcal{L}_{\mathrm{SIR\cdot SP\cdot SF}}$ achieves the highest $\mathrm{FW\text{-}SAR}$ and $\mathrm{FW\text{-}SDR}$ scores. Finally, $\mathrm{STOI}$ decreases for all weighted losses under $\mathrm{SSN}$, particularly for $\mathcal{L}_{\mathrm{logSIR}}$ and $\mathcal{L}_{\mathrm{SIR\cdot SP}}$, while the drop remains more limited for $\mathcal{L}_{\mathrm{SIR\cdot SP\cdot SF}}$.

Overall, these results indicate that weighting strategies combining interference awareness with additional spectral or temporal structure—either through learned frequency profiles or the spectral-flux–augmented scheme—yield more consistent interference suppression across noise types, whereas relying solely on local $\mathrm{SIR}$-based time–frequency modulation, such as $\mathcal{L}_{\mathrm{logSIR}}$ and $\mathcal{L}_{\mathrm{SIR\cdot SP}}$, exhibits less stable performance, particularly under spectrally shaped noise.

%Overall, these results indicate that weighting strategies combining interference awareness with additional spectral or temporal structure—either through learned frequency profiles or the spectral-flux–augmented scheme provide a more robust behavior across noise types, whereas relying solely on local $\mathrm{SIR}$-based time–frequency modulation, such as $\mathcal{L}_{\mathrm{logSIR}}$ and $\mathcal{L}_{\mathrm{SIR\cdot SP}}$, exhibit less stable performance, particularly under spectrally shaped noise.

% Unlike the $\mathrm{WN}$ condition, introducing $\mathrm{SIR}$-based weighting or learning frequency weights under $\mathrm{SSN}$ results in more noticeable variations in $\mathrm{SAR}$, $\mathrm{SDR}$, $\mathrm{FW\text{-}SDR}$, and $\mathrm{STOI}$. 

We now examine speech recognition results in terms of $\mathrm{WER}$, as shown in Fig.~\ref{fig:wer_snr}, focusing on the three proposed weighting schemes. For $\mathrm{WN}$, despite being relatively high, the gap between input and output $\mathrm{WER}$ is consistent with observations reported when comparing speech recognition-based evaluation with listening tests~\cite{drelingyte2025phoneme}. All weighted losses yield a lower $\mathrm{WER}$ than the baseline across most $\mathrm{SIR}$ levels. While $\mathcal{L}_{\mathrm{learn}}$ yields consistent improvements, both $\mathcal{L}_{\mathrm{SIR\cdot SP}}$ and $\mathcal{L}_{\mathrm{SIR\cdot SP\cdot SF}}$ significantly decrease the $\mathrm{WER}$, particularly at mid-to-high $\mathrm{SIR}$ levels. In the $\mathrm{SSN}$ condition, trends are less consistent. The loss $\mathcal{L}_{\mathrm{learn}}$ performs worse than the baseline across all $\mathrm{SIR}$ levels. Below 0~dB, the baseline achieves the lowest $\mathrm{WER}$, whereas above 0~dB both $\mathrm{SIR}$-based losses perform similarly to the baseline, with a slight advantage for $\mathcal{L}_{\mathrm{SIR\cdot SP\cdot SF}}$ at mid-to-high $\mathrm{SIR}$ levels. Overall, these results confirm previous observations: interference-aware weighting improves recognition performance, particularly under $\mathrm{WN}$ conditions. However, incorporating additional spectral or temporal structure, as in $\mathcal{L}_{\mathrm{SIR\cdot SP\cdot SF}}$, provides the most robust and consistent behavior.

\subsection{Performance on consonants and vowels}
% \definecolor{myblue}{RGB}{0,114,178}
\definecolor{myblue}{HTML}{0057B8}

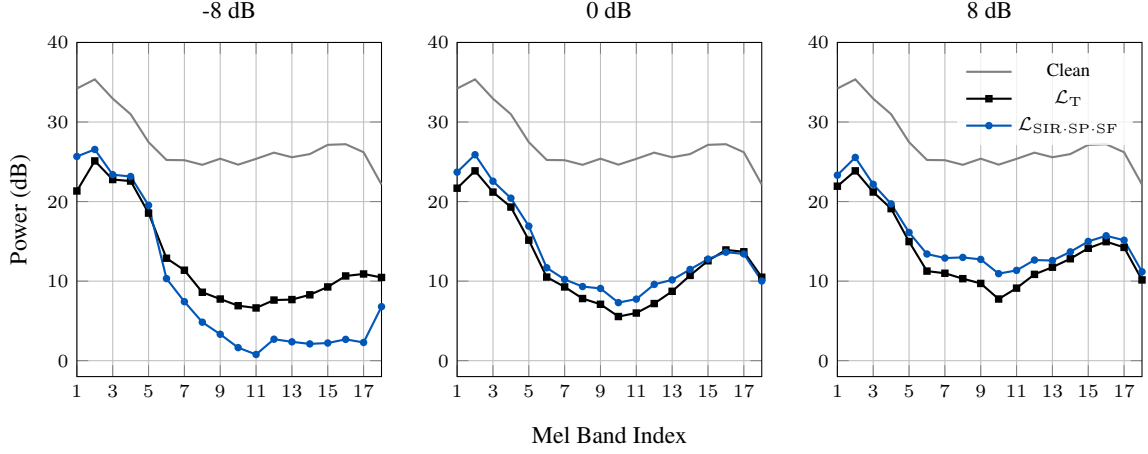
\begin{figure*}[t]
\centering
\begin{tikzpicture}
\begin{groupplot}[
    group style={
        group size=3 by 1,
        horizontal sep=1cm,
        vertical sep=1cm
    },
    width=0.33\textwidth,
    height=6cm,
    xlabel style={yshift=-5pt},
    xmin=1, xmax=18,
    xtick={1,3,5,7,9,11,13,15,17},
    ymin=-2, ymax=40,
    grid=major,
    tick label style={font=\scriptsize},
    title style={font=\small}
]

% ================= -8 dB =================
\nextgroupplot[ylabel={Power (dB)}, title={-8 dB}]
\addplot[gray, thick, mark=none] coordinates {
(1,34.21) (2,35.36) (3,32.94) (4,30.98) (5,27.47) (6,25.23)
(7,25.20) (8,24.62) (9,25.39) (10,24.64) (11,25.37) (12,26.15)
(13,25.57) (14,25.97) (15,27.13) (16,27.21) (17,26.20) (18,22.14)
};
\addplot[black, thick, mark=square*, mark size=1pt] coordinates {
(1,21.33) (2,25.11) (3,22.77) (4,22.58) (5,18.55) (6,12.88)
(7,11.37) (8,8.60) (9,7.75) (10,6.91) (11,6.63) (12,7.62)
(13,7.68) (14,8.28) (15,9.27) (16,10.65) (17,10.90) (18,10.46)
};
\addplot[myblue, thick, mark=*, mark size=1pt] coordinates {
(1,25.67) (2,26.55) (3,23.38) (4,23.15) (5,19.53) (6,10.32)
(7,7.42) (8,4.84) (9,3.32) (10,1.64) (11,0.78) (12,2.70)
(13,2.37) (14,2.11) (15,2.22) (16,2.68) (17,2.29) (18,6.79)
};

% ================= 0 dB =================
\nextgroupplot[title={0 dB}, xlabel={Mel Band Index}]
\addplot[gray, thick, mark=none] coordinates {
(1,34.21) (2,35.36) (3,32.94) (4,30.98) (5,27.47) (6,25.23)
(7,25.20) (8,24.62) (9,25.39) (10,24.64) (11,25.37) (12,26.15)
(13,25.57) (14,25.97) (15,27.13) (16,27.21) (17,26.20) (18,22.14)
};
\addplot[black, thick, mark=square*, mark size=1pt] coordinates {
(1,21.68) (2,23.86) (3,21.19) (4,19.31) (5,15.15) (6,10.51)
(7,9.27) (8,7.81) (9,7.09) (10,5.55) (11,5.99) (12,7.18)
(13,8.72) (14,10.72) (15,12.56) (16,13.92) (17,13.69) (18,10.49)
};
\addplot[myblue, thick, mark=*, mark size=1pt] coordinates {
(1,23.70) (2,25.89) (3,22.56) (4,20.43) (5,16.91) (6,11.68)
(7,10.21) (8,9.32) (9,9.08) (10,7.30) (11,7.74) (12,9.59)
(13,10.16) (14,11.47) (15,12.77) (16,13.64) (17,13.42) (18,10.02)
};

% ================= 8 dB =================
\nextgroupplot[
    title={8 dB},
    legend columns=1,
    legend style={
        draw=none,
        at={(0.97,0.97)},
        anchor=north east,
        fill=white,
        fill opacity=0.8,
        text opacity=1,
        font=\scriptsize
    }
]
\addplot[gray, thick, mark=none] coordinates {
(1,34.21) (2,35.36) (3,32.94) (4,30.98) (5,27.47) (6,25.23)
(7,25.20) (8,24.62) (9,25.39) (10,24.64) (11,25.37) (12,26.15)
(13,25.57) (14,25.97) (15,27.13) (16,27.21) (17,26.20) (18,22.14)
}; \addlegendentry{Clean}

\addplot[black, thick, mark=square*, mark size=1pt] coordinates {
(1,21.93) (2,23.86) (3,21.20) (4,19.11) (5,14.97) (6,11.26)
(7,10.99) (8,10.31) (9,9.71) (10,7.75) (11,9.12) (12,10.85)
(13,11.75) (14,12.81) (15,14.13) (16,14.96) (17,14.24) (18,10.15)
}; \addlegendentry{$\mathcal{L}_{\mathrm{T}}$}

\addplot[myblue, thick, mark=*, mark size=1pt] coordinates {
(1,23.32) (2,25.55) (3,22.17) (4,19.71) (5,16.12) (6,13.41)
(7,12.90) (8,12.98) (9,12.73) (10,10.94) (11,11.35) (12,12.65)
(13,12.59) (14,13.68) (15,14.99) (16,15.70) (17,15.14) (18,11.18)
}; \addlegendentry{$\mathcal{L}_{\mathrm{SIR}\cdot\mathrm{SP}\cdot\mathrm{SF}}$}

\end{groupplot}
\end{tikzpicture}
\caption{Comparison of estimated plosive spectra across $\mathrm{-8}$~dB, $\mathrm{0}$~dB, and $\mathrm{8}$~dB input $\mathrm{SIR}$ under $\mathrm{WN}$. The plots illustrate the spectral reconstruction performance of $\mathcal{L}_{\mathrm{T}}$ and $\mathcal{L}_{\mathrm{SIR}\cdot\mathrm{SP}\cdot\mathrm{SF}}$ relative to the clean reference spectrum.}
\label{fig:plosive_spectra}
\end{figure*}

We now report the various performance metrics computed for phoneme categories in Table~\ref{tab:phoneme_results_compact}. For clarity, we focus on comparing the proposed $\mathcal{L}_{\mathrm{SIR\cdot SP\cdot SF}}$ loss against the $\mathcal{L}_{\mathrm{T}}$ baseline. 

For consonants under $\mathrm{WN}$, $\mathcal{L}_{\mathrm{SIR\cdot SP\cdot SF}}$ improves interference suppression, as reflected by higher $\mathrm{SIR}$ and $\mathrm{FW\text{-}SIR}$ compared to the baseline, while maintaining comparable $\mathrm{FW\text{-}SDR}$ and improving $\mathrm{PA}$. Although $\mathrm{SAR}$ and $\mathrm{SDR}$ slightly decrease relative to the baseline, the drop remains limited, indicating that improved interference suppression does not induce substantial additional distortion. Under $\mathrm{SSN}$, interference reduction in terms of $\mathrm{SIR}$ and $\mathrm{FW\text{-}SIR}$ remains comparable to the baseline. Slight reductions are observed in artifact-related and overall distortion metrics, both classical and frequency-weighted. Nevertheless, $\mathcal{L}_{\mathrm{SIR\cdot SP\cdot SF}}$ results in an improvement in $\mathrm{PA}$.

For vowels under $\mathrm{WN}$, $\mathcal{L}_{\mathrm{SIR\cdot SP\cdot SF}}$ increases both $\mathrm{SIR}$ and $\mathrm{FW\text{-}SIR}$, while showing slight reductions in artifact- and distortion-related metrics. Despite these minor decreases, $\mathrm{PA}$ improves compared to the baseline, indicating that the enhanced interference suppression benefits vowel recognition. Under $\mathrm{SSN}$, $\mathcal{L}_{\mathrm{SIR\cdot SP\cdot SF}}$ achieves higher $\mathrm{SIR}$ and frequency-weighted metrics than the baseline, together with the highest vowel $\mathrm{PA}$. Despite moderate variations in $\mathrm{SAR}$ and $\mathrm{SDR}$, the overall performance remains stable.

These phoneme-level results are consistent with the utterance-level findings: $\mathcal{L}_{\mathrm{SIR\cdot SP\cdot SF}}$ tends to provide more consistent improvements across phoneme classes, particularly under $\mathrm{WN}$ conditions.

\subsection{Plosive accuracy across $\mathrm{SIR}$ levels}

\definecolor{MyBlue}{HTML}{0057B8}
\definecolor{darkred}{HTML}{B22222}

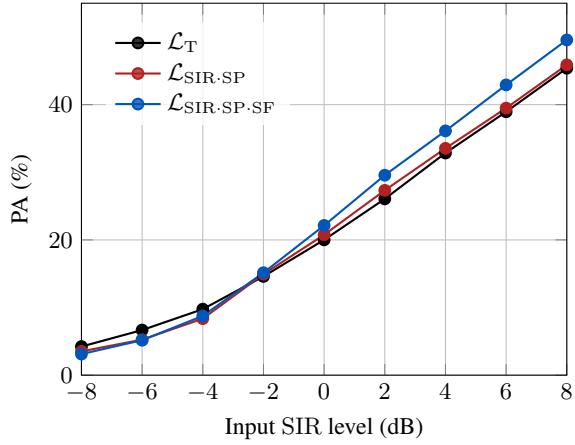
\begin{figure}[t]
\centering
\begin{tikzpicture}
\begin{axis}[
    width=\linewidth,
    height=6.5cm,
    xlabel={Input $\mathrm{SIR}$ level (dB)},
    ylabel={PA (\%)},
    xmin=-8, xmax=8,
    xtick={-8,-6,-4,-2,0,2,4,6,8},
    ymin=0, ymax=55,
    grid=major,
    % Legend moved inside and border removed
    legend style={
        at={(0.05,0.95)}, 
        anchor=north west, 
        cells={anchor=west},
        font=\small,
        draw=none,           % This removes the black border line
        fill=white,          % Keeps background to mask grid lines
        fill opacity=0.8,    % Softens the background
        text opacity=1
    },
]

% Baseline
\addplot[
    black,
    thick,
    mark=*,
    mark size=2pt
]
coordinates {
(-8,4.23) (-6,6.67) (-4,9.76) (-2,14.60) (0,20.01) (2,26.07) (4,32.85) (6,38.95) (8,45.39)
};
% SIR-SP
\addplot[
    darkred,
    thick,
    mark=*,
    mark size=2pt
]
coordinates {
(-8,3.52)(-6,5.25)(-4,8.36)(-2,14.93)(0,20.76)(2,27.33)(4,33.54)(6,39.49)(8,45.88)
};

% SIR-SP-SF
\addplot[
    MyBlue,
    thick,
    mark=*,
    mark size=2pt
]
coordinates {
(-8,3.12) (-6,5.17) (-4,8.76) (-2,15.15) (0,22.14) (2,29.57) (4,36.12) (6,42.92) (8,49.57)
};

\legend{
$\mathcal{L}_{\mathrm{T}}$,
$\mathcal{L}_{\mathrm{SIR}\cdot\mathrm{SP}}$,
$\mathcal{L}_{\mathrm{SIR}\cdot\mathrm{SP}\cdot\mathrm{SF}}$
}

\end{axis}
\end{tikzpicture}
\caption{Plosive $\mathrm{PA}$ under $\mathrm{WN}$ across input $\mathrm{SIR}$ levels.}
\label{fig:plosive_accuracy_wn}
\end{figure}

To further analyze phoneme-level performance observed in Table~\ref{tab:phoneme_results_compact}, we examine plosive recognition across $\mathrm{SIR}$ levels under $\mathrm{WN}$ (Fig.~\ref{fig:plosive_accuracy_wn}). Plosives are characterized by brief transients and rapid spectral changes, making them particularly sensitive to masking and well suited to assess the proposed spectral-flux-based weighting. %This complements the averaged results in Table~\ref{tab:phoneme_results_compact} by examining the behavior across noise levels.

Across increasing $\mathrm{SIR}$ levels, all losses exhibit a monotonic improvement in plosive $\mathrm{PA}$. At very low $\mathrm{SIR}$, performance remains similar, with the baseline slightly higher. However, from mid $\mathrm{SIR}$ levels onward, $\mathcal{L}_{\mathrm{SIR\cdot SP\cdot SF}}$ significantly outperforms the baseline from 0~dB and $\mathcal{L}_{\mathrm{SIR\cdot SP}}$ from 2~dB onward, with the performance gap widening at higher $\mathrm{SIR}$ levels.

This trend confirms that incorporating $\mathrm{SIR}$-aware and spectral-flux-based weighting enhances recognition of transient phonemes such as plosives, particularly in moderate-to-high $\mathrm{SIR}$ conditions where interference suppression becomes more effective.

\subsection{Spectral analysis of plosives}

To better understand the spectral behavior underlying the plosive results, Fig.~\ref{fig:plosive_spectra} compares the estimated spectra to the clean reference across $\mathrm{-8}$, $\mathrm{0}$, and $\mathrm{8}$~dB under $\mathrm{WN}$ in the Mel scale.

At $\mathrm{-8}$~dB, both methods remain far from the clean spectrum. Up to band~6, the two estimates are similar. Beyond band~6, $\mathcal{L}_{\mathrm{T}}$ remains closer to the clean reference, while $\mathcal{L}_{\mathrm{SIR\cdot SP\cdot SF}}$ shows larger deviation in the mid-to-high bands.
At $\mathrm{0}$~dB, $\mathcal{L}_{\mathrm{SIR\cdot SP\cdot SF}}$ becomes slightly closer to the clean spectrum than the baseline, particularly across bands~8–13, where it better tracks the clean spectral shape. At $\mathrm{8}$~dB, the same tendency is observed with a larger margin, especially over bands~6–12, where $\mathcal{L}_{\mathrm{SIR\cdot SP\cdot SF}}$ more closely follows the clean spectrum.

%Overall, when the input $\mathrm{SIR}$ level increases the benefit of $\mathcal{L}_{\mathrm{SIR\cdot SP\cdot SF}}$ in reconstructing the mid-frequency structure of plosives becomes more pronounced. 

Overall, the benefit of $\mathcal{L}_{\mathrm{SIR\cdot SP\cdot SF}}$ in reconstructing the mid-frequency structure of plosives is mainly observed at moderate and high input $\mathrm{SIR}$ levels. At very low $\mathrm{SIR}$, the baseline remains closer to the clean spectrum in several mid-to-high frequency bands. These less adverse listening conditions are particularly relevant for hearing-aid applications.

\section{Conclusion}

We proposed a TF weighting framework for $\mathrm{SDR}$-based speech enhancement that explicitly models speech presence, local $\mathrm{SIR}$, and transient dynamics. The proposed formulations reshape the contribution of TF bins according to speech--noise competition, while preserving the stability of $\mathrm{SDR}$ optimization. Experimental results show that incorporating $\mathrm{SIR}$-aware and spectral-flux-based modulation improves interference suppression and phoneme-level reconstruction.
%Experimental results show that incorporating $\mathrm{SIR}$-aware and spectral-flux-based modulation improves interference suppression and phoneme-level reconstruction, particularly for consonants and plosives under WN and moderate $\mathrm{SIR}$ conditions. These improvements are reflected not only in frequency-weighted objective metrics but also in terms of speech recognition.
Future work will explore additional perceptually-based phenomena, such as frequency-dependent speech reception thresholds, band-specific masking, or intelligibility-weighted filters. Such directions may better align training with phoneme-level perception and hearing-assistive objectives.

\newpage
\section{Acknowledgments}
This research was supported by the French National Research Agency as part of the REFINED project, “REal-time artiFicial INtelligence for hEaring aiDs” (ANR-21-CE19-0043). Experiments presented in this paper were carried out using the \href{https://www.grid5000.fr}{Grid'5000} testbed, supported by a scientific interest group hosted by Inria and including CNRS, RENATER and several Universities as well as other organizations.

\section{Generative AI Use Disclosure}
We acknowledge the ISCA policy regarding the use of generative AI tools. The authors declare that generative AI tools have been used solely to correct grammar. No such tools were used to write significant parts of this manuscript.

\bibliographystyle{IEEEtran}
\bibliography{mybib}

\end{document}